\documentclass[aps,prl,twocolumn,showpacs,amsmath,amssymb,superscriptaddress]{revtex4}
\usepackage{graphicx}
\usepackage{dcolumn}
\usepackage{bm}
\usepackage{natbib}
\usepackage{amsmath}
\usepackage{array}

\begin{document}
\title{Microscopic picture of aging in SiO$_2$}

\author{Katharina Vollmayr-Lee}
 \email{kvollmay@bucknell.edu}
\affiliation{Department of Physics and Astronomy, Bucknell University,
      Lewisburg, Pennsylvania 17837, USA}
\author{Robin Bjorkquist}
\affiliation{Department of Physics, Cornell University, Ithaca, New York 14853, USA}
\author{Landon M.\ Chambers}
\affiliation{Department of Physics, Texas A\&M University, 
      College Station, TX 77843, USA}

\date{September 18, 2012}
\begin{abstract}
We investigate the aging dynamics of amorphous SiO$_2$ 
via molecular dynamics simulations of a quench from a high temperature
$T_{\rm i}$ to a lower temperature $T_{\rm f}$. 
We obtain a microscopic picture of aging dynamics by
analyzing single particle trajectories, identifying jump events 
when a particle escapes the cage formed by its neighbors, 
and by determining how these jumps depend 
on the waiting time $t_{\rm w}$, 
the time elapsed since the temperature quench to $T_{\rm f}$. 
We find that the only $t_{\rm w}$-dependent microscopic quantity is
the number of jumping particles per unit time, which decreases with 
age.  Similar to previous studies for fragile glass formers, 
we show here for the strong glass former SiO$_2$ that
neither the distribution
of jump lengths nor the distribution of times spent 
in the cage are $t_{\rm w}$-dependent. 
We conclude that the microscopic aging dynamics is surprisingly
similar for fragile and strong glass formers.

\end{abstract}

%
\maketitle
If a system is quenched from a high temperature
$T_{\rm i}$ to a lower temperature $T_{\rm f}$ below the glass transition, 
crystallization is avoided and
a glass is formed. The resulting out of equilibrium 
(aging) dynamics has been hotly debated for the last decades and 
remains unclear \cite{glassbook,berthierbiroli2011}.
Most previous studies on the aging dynamics investigated
quantities which are averages over all particles in the 
system, such as 
mean squared displacement, incoherent
intermediate scattering function,
dynamic susceptibility, 
and energy 
\cite{colin2011,HeuerJPCM2008,MasriPRE2010,parsaeian09,Rehwald2010,kvl2010,WarrenPRE2008}. 
On the other hand much less is known about single particle 
dynamics 
during aging. 
For colloids, Cianci et al.\ 
investigated the structure 
\cite{CianciSolidState2006,CianciAIPConfProc2006} 
and Yunker et al.\ 
\cite{YunkerYodh2009}
focused on irreversible rearrangements 
as function of waiting time $t_{\rm w}$.
Warren and Rottler used computer simulations to 
investigate single particle hopping events  for a
binary Lennard-Jones mixture without shear as well as for
polymers with and without shear 
\cite{WarrenEPL2009,WarrenPRL2010,WarrenJCP2010}.
To gain a more complete picture of the microscopic processes during aging,
we study single particle hopping (jump) events  
for the very different  glass former SiO$_2$.
Whereas the systems of Warren and Rottler are fragile glass
formers, SiO$_2$ belongs to the class of strong glass 
formers \cite{glassbook}.\\
We determine the number of jumping particles per unit time, 
the jump length, and the time spent in a cage 
for a wide range of waiting times $t_{\rm w}$ and 
for several choices of $T_{\rm i}$ and $T_{\rm f}$.
   To study the aging dynamics of amorphous silica we 
carried out molecular dynamics (MD) simulations using the
BKS potential \cite{beest_90} for the particle interactions.
Starting from 20 independent
fully equilibrated configurations at high temperatures $T_{\rm
i} \in \{5000$\,K$, 3760$\,K\}, the system is quenched instantaneously
to lower
temperatures $T_{\rm f} \in \{ 2500$\,K, $2750$\,K, $3000$\,K, $3250$\,K$\}$.
To keep the temperature at $T_{\rm f}$ constant and
to disturb the dynamics minimally, 
the Nos\'e-Hoover thermostat 
was applied only
for the first $0.33$~ns (NVT), and the simulation was continued 
in the NVE ensemble for $33$~ns during which $T_{\rm f}$ stayed
constant. 
For more information on details of the simulation see \cite{kvl2010}.

\begin{figure}
\includegraphics[width=3.0in]{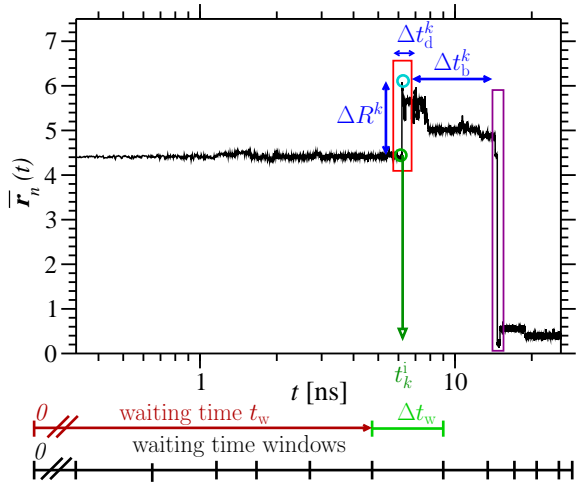}
\caption{\sf (Color online) As an example for the time-averaged trajectory 
$\overline{\mathbf r_n}(t_l)$ we show here the z-component 
$\overline{\mathbf r_{n,z}}$ 
 for the oxygen atom $n=315$ for a single simulation run at $T_f=2500$~K 
which had been quenched from $T_i=3760$~K. For clarity, only 
a fraction of the simulation time is shown.
}
\label{fig:jumpdef}
\end{figure}

We focus on the microscopic dynamics 
at the lower temperature $T_{\rm f}$ by
analyzing the single particle trajectories $\mathbf r_n(t)$.
During the 
production runs 
at $T_{\rm f}$ we stored 
average positions $\overline{\mathbf r}_n(t_l)$ and fluctuations
$\sigma_n(t_l)=\sqrt{\overline{{\mathbf r}_n^2}(t_l)
       -\left (\overline{\mathbf r}_n(t_l) \right )^2}$ 
for each particle $n$ 
at times $t_l= l \times (0.00327$~ns$)$. 
Here $\overline{(\ldots)}$ correspond 
to averages over 3200 MD steps and 2000 MD steps for the 
NVT and NVE simulation runs respectively.
We then use the resulting $\overline{\mathbf r}_n(t_l)$ 
to identify jump events. For example Fig.~\ref{fig:jumpdef} shows
the y-component of $\overline{\mathbf r}_n(t_l)$
for $n=315$; rectangular boxes indicate identified jumps.
We define a particle $n$ to undergo a jump if its change in average 
position 
\begin{equation}
\Delta \overline{r}_n 
       = \left | \overline{\mathbf r}_n(t_l) - \overline{\mathbf r}_n(t_{l-4}) \right |
\end{equation}
satisfies 
\begin{equation}
\Delta \overline{r}_n > 3 \sigma_{\alpha}
\label{eq:jumpdef}
\end{equation}
where
$\sigma_{\alpha}$ is the average fluctuation  size for particle 
type $\alpha \in \{\mbox{Si,O}\}$. Since $\sigma_{\alpha}$ is 
intended to be a measure of average fluctuations during each particles
rattling within its cage of neighbors, we first determine the 
estimate $\sigma^2_{{\rm est},\alpha}$ by averaging 
$\left (\sigma_n(t_l)\right )^2$ over all times $t_l$ 
of a given simulation run at $T_{\rm f}$
and over all particles of the same type $\alpha$. We then determine
$\sigma_{\alpha}$ by redoing the
average over $\left (\sigma_n(t_l)\right )^2$, 
but by averaging only over times for which 
$\left (\sigma_n(t_l)\right )^2 < 3 \sigma^2_{{\rm est},\alpha}$
which roughly excludes jumps from the average.
Note that the definition of Eq.~(\ref{eq:jumpdef})
is similar, but not identical to 
our analysis in \cite{vollmayr04,vollmayr06}.
To verify that our results are independent of the 
details of the jump definition, we replaced Eq.~(\ref{eq:jumpdef})
with $\Delta \overline{r}_n > \sqrt{2} \sigma_{\alpha}$ 
and found indeed qualitatively the same results as they are presented
here, for which we used Eq.~(\ref{eq:jumpdef}).

We thus identify for all simulation runs 
all jump events occurring during 
the production run at $T_{\rm f}$.
For each 
jump event $k$ we determine the particle $n_k$
jumping from average position 
$\left (\overline{\mathbf r_{n_k}} \right )^{\rm i}$
at time $t_k^{\rm i}$ 
to average position 
$\left (\overline{\mathbf r_{n_k}} \right )^{\rm f}$
at time $t_k^{\rm f}$ 
(see in Fig.~\ref{fig:jumpdef} dark green and cyan circles).

Our focus is on the dynamics of the system 
as it is aging over time. We investigate it via the jump events
and their dependence on the waiting time $t_{\rm w}$, 
i.e.\ the time elapsed since the temperature quench to $T_{\rm f}$. 
We divide the simulation run into waiting time windows,
as indicated in Fig.~\ref{fig:jumpdef} \footnote{In 
simulation time units ($1.0217 \times 10^{-5}$~ns) we
used the borders $0, \large ( 1000 \times 2^{m_1}$ for
$m_1=0,1,\ldots,6 \large ), \large (64000+49500 \times 2^{m_2}$ for
$m_2=0,\ldots,3 \large ), \large (64000+m_3 \times 396000$ for
$m_3=2,\ldots,8 \large )$.}. 
For each jump event $k$ with jump time 
$t_k^{\rm i}$ we determine 
the waiting time window which includes $t_k^{\rm i}$
(in Fig.~\ref{fig:jumpdef} the light green waiting time window)
and assign to this waiting time window the
waiting time $t_{\rm w}$ of the left border of the 
selected time window (in Fig.~\ref{fig:jumpdef} red arrow).

\begin{figure}
\includegraphics[width=3.0in]{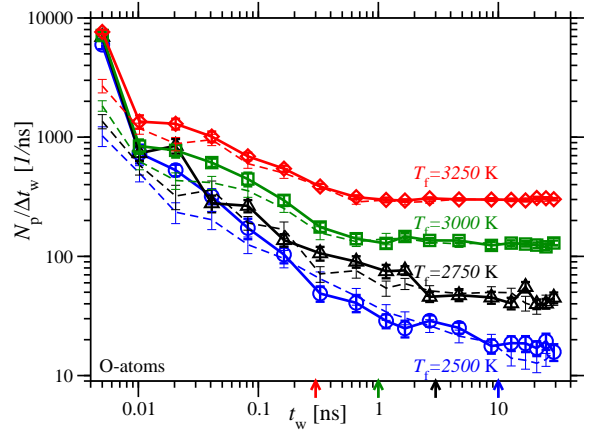}
\caption{\sf
(Color online) Number of jumping particles $N_{\rm p}$ 
per time $\Delta t_{\rm w}$
as function of waiting time $t_{\rm w}$ for the case 
of O-atoms and $T_{\rm i}=5000$~K (bold lines and symbols)
and $T_{\rm i}=3760$~K (dashed thin lines).
To be able to include on the logarithmic scale the data-point for 
the first time window at $t_{\rm w}=0$, we plot 
$\frac{N_{\rm p}}{\Delta t_{\rm w}}(t_{\rm w}=0)$ 
instead at $t_{\rm w}=0.005$~ns.
For comparison the arrows indicate
the equilibrium times $t_{\rm eq}^C$ ($t_{23}$ in \cite{kvl2010}).
}
\label{fig:nojp}
\end{figure}

We therefore obtain jump statistics for each waiting time window starting
at time $t_{\rm w}$ and of duration $\Delta t_{\rm w}$ 
(see Fig.~\ref{fig:jumpdef}).
In Fig.~\ref{fig:nojp} we show the number of distinct particles jumping 
per observation time $\Delta t_{\rm w}$ 
as function of waiting time $t_{\rm w}$
\footnote{To avoid that all particles jump, we choose a small 
enough window. For the case of $\Delta t_{\rm w} > 0.506$~ns
we therefore divide the waiting time window into 
subwindows of size $\Delta t = 0.506$~ns and average over
$\frac{N_{\rm p}}{\Delta t}$.}. 
We find for all investigated $T_{\rm f}$ and both $T_{\rm i}$ 
a clear $t_{\rm w}$-dependence. 
With increasing waiting time $\frac{N_{\rm p}}{\Delta t_{\rm w}}$ decreases 
following roughly a power law  until equilibrium is 
reached and $\frac{N_{\rm p}}{\Delta t_{\rm w}}(t_{\rm w})$ 
becomes independent of $t_{\rm w}$ and $T_{\rm i}$. 
The power law exponents are approximately the same for O- and Si-atoms
in the range $[-0.6/$ns$,-0.3/$ns$]$.
As one might expect, the larger $T_{\rm f}$ the more particles jump
and the earlier the equilibrium time $t_{\rm eq}^{\rm j}$, i.e.\ 
the time when $\frac{N_{\rm p}}{\Delta t_{\rm w}}$ levels off. 
For comparison we include in Fig.~\ref{fig:nojp} 
the equilibrium times $t_{\rm eq}^C$ determined 
via the intermediate incoherent
scattering function $C_q(t_{\rm w},t_{\rm w}+t)$ ($t_{\rm eq}^C=t_{23}$ in 
\cite{kvl2010}). 
We find 
 $t_{\rm eq}^{\rm j} \approx t_{\rm eq}^C$,
i.e.\ agreement between 
the {\em microscopic}
equilibrium time  $t_{\rm eq}^{\rm j}$ (single particle jumps)
and the {\em macroscopic} equilibrium $t_{\rm eq}^C$ 
($C_q$ includes a particle average).

\begin{figure}
\includegraphics[width=3.0in]{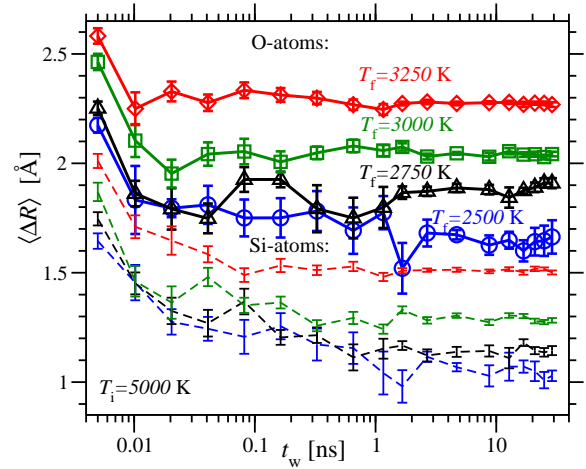}
\caption{\sf
(Color online) Jump length $\langle \Delta R \rangle$ 
(see Eq.~(\ref{eq:dRif}) and Fig.~\ref{fig:jumpdef})
as function of waiting time $t_{\rm w}$ for the case 
of $T_{\rm i}=5000$~K and O-atoms (bold lines and symbols)
and Si-atoms (dashed thin lines).
Similar to Fig.~\ref{fig:nojp} we plot 
$\langle \Delta R \rangle (t_{\rm w}=0)$
at $t_{\rm w}=0.005$~ns.
}
\label{fig:dRif}
\end{figure}

\begin{figure}
\includegraphics[width=3.0in]{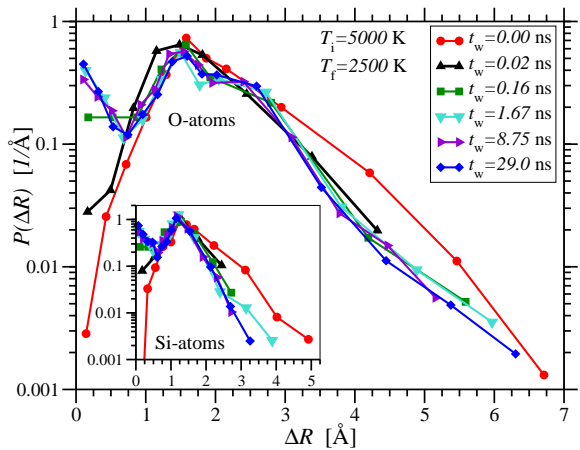}
\caption{\sf
(Color online) Distribution of the jump length
$P(\Delta R)$ 
for the case of $T_{\rm i}=5000$~K, $T_{\rm f}=2500$~K and for 
O-atoms and in the inset for Si-atoms. Different colors indicate
waiting time $t_{\rm w}$.
}
\label{fig:PofdRif}
\end{figure}

Next we test whether the $t_{\rm w}$-dependence 
manifests itself also in a microscopic length scale.
As sketched in Fig.~\ref{fig:jumpdef}, we define the 
jump length of event $k$ of particle $n_k$ jumping 
at time $t_k^{\rm i}$ from
$\left (\overline{\mathbf r_{n_k}} \right )^{\rm i}$ to 
$\left (\overline{\mathbf r_{n_k}} \right )^{\rm f}$
to be
\begin{equation}
\Delta R^k = \left | \left (\overline{\mathbf r_{n_k}} \right )^{\rm f}
                      - \left (\overline{\mathbf r_{n_k}} \right )^{\rm i}
                   \right |
          \qquad \mbox{.}
\label{eq:dRif}
\end{equation}
Similar to above, we investigate the $t_{\rm w}$-dependence of
$\langle \Delta R \rangle$ by 
including in the average only events for which 
$t_k^{\rm i}$ belong to the same waiting time window.
The resulting Fig.~\ref{fig:dRif} shows  
that $\langle \Delta R \rangle$ 
for oxygen atoms (solid thick lines with symbols)
is independent of $t_{\rm w}$ 
(with the only exception of the first time-window),
and for silicon atoms (dashed thin lines) $\langle \Delta R \rangle$
is only slightly $t_{\rm w}$-dependent.
This is in stark contrast to $\frac{N_{\rm p}}{\Delta t_{\rm w}}$ 
of Fig.~\ref{fig:nojp}, which shows strong $t_{\rm w}$-dependence. 
The $t_{\rm w}$-independence of 
$\Delta R$ holds 
true even for the distribution 
$P(\Delta R)$,
both for O- and for Si-atoms, 
as shown in Fig.~\ref{fig:PofdRif} for the case of
$T_{\rm i}=5000$~K, $T_{\rm f}=2500$~K.
We find similar results for all other investigated 
$T_{\rm i}$ and $T_{\rm f}$. 
Consistent with Fig.~\ref{fig:dRif}, we find only $t_{\rm w}$-dependence 
for $t_{\rm w} \lessapprox 0.02$~ns (which corresponds in an 
experiment to the undetectable instant of an infinitely fast
quench).
For $t_{\rm w} > 0.02$ an additional peak occurs
at $\Delta R \approx 0$  which is mostly due to reversible jumps
(as defined in \cite{vollmayr04}).
Furthermore we find exponential tails 
$P(\Delta R) \sim \exp \left( - \Delta R/R_{\rm decay} \right )$ 
with $R_{\rm decay} \approx 0.8$ and $0.3 \,\AA$ for O- and Si-atoms 
respectively (similar to the results for a binary Lennard Jones mixture 
\cite{WarrenEPL2009}).

\begin{figure}
\includegraphics[width=3.0in]{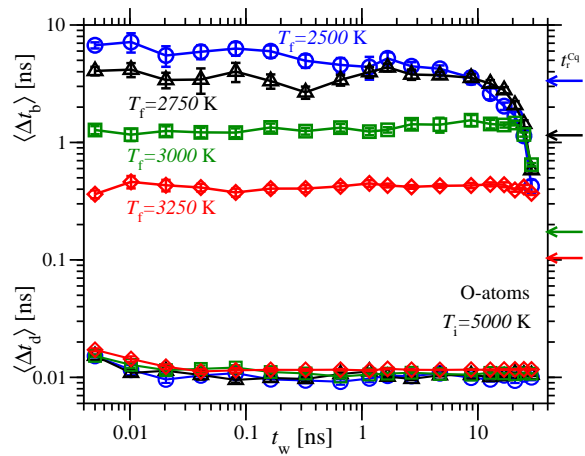}
\caption{\sf
(Color online) We show here 
average jump duration $\langle \Delta t_{\rm d}\rangle$ 
(lower four curves) and time between 
successive jumps of the same particle $\langle \Delta t_{\rm b}\rangle$ 
(top four curves)
using the definitions of Eq.~(\ref{eq:dtif}) and Eq.~(\ref{eq:dt34}) 
and Fig.~\ref{fig:jumpdef}.
The arrows on the right indicate 
$t_{\rm r}^{\rm Cq}(t_{\rm w}=23.98$~ns$)$ of \cite{kvl2010}.
We include $\Delta t_{\rm d}(0$~ns$)$ and $\Delta t_{\rm b}(0$~ns$)$
at $t_{\rm w}=0.005$~ns.
}
\label{fig:dt}
\end{figure}

\begin{figure}[h]
\includegraphics[width=3.0in]{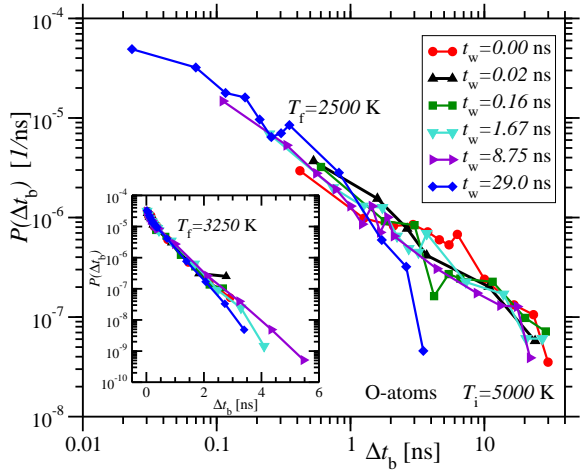}
\caption{\sf
(Color online) Distribution of times between jumps $P(\Delta t_{\rm b})$ 
for O-atoms, $T_{\rm i}=5000$~K and for $T_{\rm f}=2500$~K and 
in the inset for $T_{\rm f}=3250$~K. Different symbols (and colors) 
correspond to different waiting times $t_{\rm w}$.
}
\label{fig:Pofdtdrtws}
\end{figure}

\begin{figure}[h]
\includegraphics[width=3.0in]{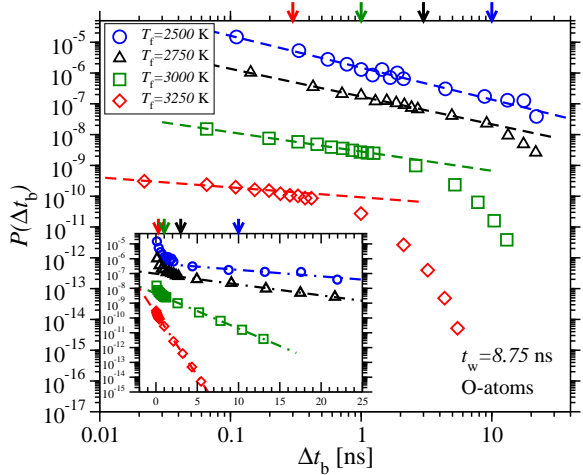}
\caption{\sf
(Color online) $P(\Delta t_{\rm b})$
for fixed $t_{\rm w}=8.75$~ns, $T_{\rm i}=5000$~K and for O-atoms
as log-log plot in the main figure and as log-lin plot in the inset.
Different symbols (and colors) correspond to different final temperature
$T_{\rm f}$. 
Dashed lines are power law fits with exponents 
$-1.0, -0.9, -0.6, -0.3$ and dot-dashed lines are exponential fits
$P(\Delta t_{\rm b}) \sim \exp \left( - \Delta t_{\rm b}/t_{\rm decay} \right )$ 
with $t_{\rm decay}=10, 6, 2, 0.5$~ns for 
$T_{\rm f}=2500, 2750, 3000, 3250$~K respectively.
As in Fig.~\ref{fig:nojp}, we include for comparison arrows which indicate
the equilibrium times $t_{\rm eq}^C$ 
\cite{kvl2010}. For clarity, $P(\Delta t_{\rm b})$ has been 
shifted by a factor of $10^{-1}$/$10^{-3}$/$10^{-5}$ for $T_{\rm f}=2750$/$3000$/$3250$~K
respectively. 
}
\label{fig:PofdtdrTs}
\end{figure}

With the conclusion from Figs.~\ref{fig:dRif} and \ref{fig:PofdRif}
that the length scale $\Delta R$ is 
$t_{\rm w}$-independent, we investigate next the 
time scales associated with the single particle jumps. 
We define the duration of a jump event $k$  to be
\begin{equation}
\Delta t_{\rm d}^k = t_k^{\rm f} - t_k^{\rm i}
\label{eq:dtif}
\end{equation}
(see Fig.~\ref{fig:jumpdef}) and the time between successive 
jumps of the same particle 
\begin{equation}
\Delta t_{\rm b}^k = t_{k+1}^{\rm i} - t_k^{\rm f}
\label{eq:dt34}
\end{equation}
that means the time spent in the cage before the same particle jumps 
again (see Fig.~\ref{fig:jumpdef}).
The resulting $\langle \Delta t_{\rm d}\rangle$ 
and $\langle \Delta t_{\rm b}\rangle$ 
are shown in Fig.~\ref{fig:dt}.
The time between jumps $\langle \Delta t_{\rm b} \rangle$ 
is several magnitudes
larger than $\langle \Delta t_{\rm d} \rangle$.
For comparison with $\langle \Delta t_{\rm b} \rangle$ we include
arrows on the right to indicate 
$t_{\rm r}^{\rm Cq}(t_{\rm w}=23.98$~ns$)$ of \cite{kvl2010}, which is
defined to be the time for which
$C_q(t_{\rm w},t_{\rm w}+t_{\rm r}^{\rm Cq})=0.625$. 
Since $\langle \Delta t_{\rm b} \rangle > t_{\rm r}^{\rm Cq}$, we conclude 
that $\langle\Delta t_{\rm b} \rangle$ is characterizing $\alpha$-relaxation.
As above, we determined the $t_{\rm w}$-dependence by 
averaging $\Delta t_{\rm d}^k$ and $\Delta t_{\rm b}^k$ 
for all jump events $k$ for which $t_k^{\rm i}$ belongs to 
the same waiting time window. 
By choosing this 
definition of $\langle \Delta t_{\rm b} \rangle$  
we prevent artifacts due to 
the different time window sizes, because only $t_k^{\rm i}$ (instead 
of $\Delta t_{\rm b}^k$) is 
required to be in the time window of consideration. 
For large $t_{\rm w}$, however,
the finite simulation run time $t_{\rm tot}=33.33$~ns, 
causes 
$\langle \Delta t_{\rm b} \rangle$ to decrease for waiting times 
$t_{\rm w} \gtrapprox (t_{\rm tot}-\Delta t_{\rm b})$.
Ignoring this $t_{\rm tot}$-specific decrease, we therefore obtain
the surprising result that $\langle \Delta t_{\rm b} \rangle$ is 
independent of $t_{\rm w}$. This independence of $t_{\rm w}$ 
holds not only for the average $\langle \Delta t_{\rm b} \rangle$, but even 
for the whole distribution $P(\Delta t_{\rm b})$, as shown 
in Fig.~\ref{fig:Pofdtdrtws}. 
Also in Fig.~\ref{fig:Pofdtdrtws} we notice that
$P(\Delta t_{\rm b}) \sim \Delta t_{\rm b}^{-1}$
at $T_{\rm f}=2500$~K, whereas 
$P(\Delta t_{\rm b}) \sim \exp \left( - \Delta t_{\rm b}/t_{\rm decay} \right )$
at $T_{\rm f}=3250$~K. 
In Fig.~\ref{fig:PofdtdrTs} we show how $P(\Delta t_{\rm b})$ 
plotted versus $\Delta t_{\rm b}$ changes with the final temperature, 
for a fixed $t_{\rm w}=8.75$~ns. We observe that
at intermediate temperatures, i.e.\ $T_{\rm f}=2750$~K and $T_{\rm f}=3000$~K,
there is a crossover from power law to exponential decay.
For comparison we include in Fig.~\ref{fig:PofdtdrTs}
the same arrows as in Fig.~\ref{fig:nojp}, which indicate
the equilibrium times $t_{\rm eq}^C$. 
The crossover time occurs approximately at the
same time when $\frac{N_{\rm p}}{\Delta t_{\rm w}}(t_{\rm w})$ 
and $C_q(t_{\rm w},t_{\rm w}+t)$ reach equilibrium.
A similar crossover has been observed for kinetically constrained 
models 
(see Fig.~10 of \cite{JungGarrahanChandlerJCP2005})
and for a binary Lennard-Jones mixture 
(see Fig.~2 of \cite{DoliwaHeuerPRL2003}).

In summary, we obtain the following microscopic picture of aging:
both the distribution
of jump length and the distribution of times spent in 
the cage $P(\Delta t_{\rm b})$ 
are independent of waiting time $t_{\rm w}$
(similar to the results of Warren and Rottler 
\cite {WarrenEPL2009,WarrenPRL2010,WarrenJCP2010}).
Instead the only $t_{\rm w}$-dependent microscopic
quantity is the number of jumping particles per time, which
decreases with increasing $t_{\rm w}$ (similar to the results of 
Yunker et al.\ \cite{YunkerYodh2009}).
This is consistent with the first hop time results reported in
\cite {WarrenEPL2009,WarrenPRL2010,WarrenJCP2010}.
In agreement with kinetically constrained models 
$P(\Delta t_{\rm b})$ shows
a crossover from power law to exponential decay 
\cite{JungGarrahanChandlerJCP2005}.
Our results for the strong glass former SiO$_2$ are 
surprisingly similar to the fragile 
glass former results \cite {WarrenEPL2009,WarrenPRL2010,WarrenJCP2010}.
\begin{acknowledgments}
RB and LMC were supported by NSF REU grants
PHY-0552790 and REU-0997424.
We thank 
A.~Zippelius and H.E.~Castillo for comments 
on an earlier version 
of this manuscript. 
KVL thanks A.~Zippelius 
and the Institute of Theoretical Physics, University of G{\"{o}}ttingen,
for hospitality and 
financial support via the SFB 602.
\end{acknowledgments}

\bibliography{SiO2jumpsPRL_v6}

\end{document}